\documentclass[runningheads]{llncs}
\usepackage[rgb,x11names]{xcolor}
\usepackage{amssymb}
\usepackage{amsmath}
\usepackage{subfig}
\usepackage{hyperref}
\usepackage{graphicx}
\usepackage[]{algorithm2e}
\usepackage{mathrsfs}
\usepackage{pgfplots}
\usepackage{tablefootnote}
\usepackage{booktabs}
\usepackage[labelfont=bf]{caption}
\usepackage{wrapfig}
\usepackage{bm}
\newsavebox\CBox
\def\textBF#1{\sbox\CBox{#1}\resizebox{\wd\CBox}{\ht\CBox}{\textbf{#1}}}

\usetikzlibrary{matrix}

\usepackage{marginnote}

\usepackage{bbding}

\usepackage{soul}

\usepackage{xcolor}
\usepackage{lipsum}
\usepackage[draft]{todonotes}   
\usepackage{pgfplots}
\pgfplotsset{width=14cm,compat=1.8}
\usepackage{pgfplotstable}
\pgfplotsset{/pgfplots/ybar legend/.style={
		/pgfplots/legend image code/.code={
			\draw[##1,/tikz/.cd,bar width=3pt,yshift=-0.2em,bar shift=0pt]
			plot coordinates {(0cm,0.8em)};},},}

\usepgfplotslibrary{groupplots}

\def\ie{\emph{i.e.}}

\makeatletter 
\newcommand\figcaption{\def\@captype{figure}\caption} 
\newcommand\tabcaption{\def\@captype{table}\caption} 
\makeatother

\usepackage{graphicx}
\begin{document}
\title{$\textup{M}^2\textup{Net}$: 
Multi-modal Multi-channel Network for Overall Survival Time Prediction of Brain Tumor Patients}
%
%
%
\authorrunning{T. Zhou \emph{et al}.}

\author{Tao Zhou\inst{1}  \and      
Huazhu Fu\inst{1}\Envelope \and     
Yu Zhang\inst{2} \and               
Changqing Zhang\inst{3} \and  \\    
Xiankai Lu\inst{1}\and              
Jianbing Shen\inst{1}\Envelope \and 
Ling Shao\inst{4,1}                  
}


\institute{$^1$ Inception Institute of Artificial Intelligence, Abu Dhabi, UAE.\\
$^2$ Department of Bioengineering, Lehigh University, Bethlehem, PA 18015, USA. \\
$^3$ College of Intelligence and Computing, Tianjin University, China.\\
$^4$ Mohamed bin Zayed University of Artificial Intelligence, Abu Dhabi, UAE. \\
{\tt huazhu.fu@inceptioniai.org, shenjianbingcg@gmail.com}}

\titlerunning{$\textup{M}^2\textup{Net}$ for Overall Survival Time Prediction}

%
%
\maketitle              
\begin{abstract}
Early and accurate prediction of overall survival (OS) time can help to obtain better treatment planning for brain tumor patients. Although many OS time prediction methods have been developed and obtain promising results, there are still several issues. \emph{First}, conventional prediction methods rely on radiomic features at the local lesion area of a magnetic resonance (MR) volume, which may not represent the full image or model complex tumor patterns. \emph{Second}, different types of scanners (\emph{i.e.}, multi-modal data) are sensitive to different brain regions, which makes it challenging to effectively exploit the complementary information across multiple modalities and also preserve the modality-specific properties. \emph{Third}, existing methods focus on prediction models, ignoring complex data-to-label relationships. To address the above issues, we propose an end-to-end OS time prediction model; namely, Multi-modal Multi-channel Network ($\textup{M}^2\textup{Net}$). Specifically, we first project the 3D MR volume onto 2D images in different directions, which reduces computational costs, while preserving important information and enabling pre-trained models to be transferred from other tasks. Then, we use a modality-specific network to extract implicit and high-level features from different MR scans. A multi-modal shared network is built to fuse these features using a bilinear pooling model, exploiting their correlations to provide complementary information. Finally, we integrate the outputs from each modality-specific network and the multi-modal shared network to generate the final prediction result. Experimental results demonstrate the superiority of our $\textup{M}^2\textup{Net}$ model over other methods.
\end{abstract}
\section{Introduction}

Brain tumors are the most common and difficult-to-treat malignant neurologic tumors, with the highest mortality rate. Gliomas account for about 70\% of primary brain tumors in adults \cite{ricard2012primary}. Gliomas can be less aggressive (\emph{i.e.}, low grade), yielding a life expectancy of several years, or more aggressive (\emph{i.e.}, high grade), with a life expectancy of at most two years \cite{havaei2017brain}. The prognosis of gliomas is often measured by the overall survival (OS) time, which varies largely across individuals. Thus, timely and accurate OS time prediction for brain tumor patients is of great clinical importance and could benefit individualized treatment care.

Recently, magnetic resonance imaging (MRI) has played an important role in the study of glioma prognosis~\cite{pope2005mr,zacharaki2012survival,razek2018differentiation,zhou2019deep,liu2019overall,fan2019birnet,zhu2020multivariate}. By using informative imaging phenotypes from multi-modal MR scans (\emph{e.g.},  native (T1), T1 contrast enhanced (T1ce), T2-weighted (T2), and Fluid Attenuated Inversion Recovery (FLAIR)), several methods have been proposed for survival prediction. For example, Pope \emph{et al.} \cite{pope2005mr} found that non-enhancing tumor and infiltration areas are helpful for OS prediction by analyzing 15 MRI features. The study in \cite{liu2017relationship} also used all kinds of informative radiomic features extracted from raw images for OS time prediction. Complementary information from multi-modal data has also played an important role in helping improve prediction performance. For example, Zhou \emph{et al.} \cite{zhou2017tpcnn} extracted informative radiomic features based on segmented tumors from multi-modal data and then used the XGBoost as classifier. Feng \emph{et al.} \cite{feng2018brain} proposed a linear model for survival prediction using imaging and non-imaging features extracted from multi-modal data. Isensee \emph{et al.} \cite{isensee2017brain} conducted OS prediction by training an ensemble of a random forest regressor. Nie \emph{et al.} \cite{nie2019multi} used a 3D deep learning model to extract multi-modal multi-channel features and then feeding them into a support vector machine (SVM) classifier for OS prediction.

Although several methods have been developed for this task, there still exist several issues. \emph{First}, conventional prediction approaches rely on radiomic features at the local lesion area of an MR volume. However, these shallow and low-level features might not fully characterize the image or model the tumor's complex patterns. \emph{Second}, most methods simply concatenate features from different modalities into a single feature vector without considering the underlying correlations across multi-modal data. Moreover, rare methods exploit the correlations while also preserving the modality-specific attributes. However, both of aspects are important for improving model performance in multi-modal/view learning \cite{zhang2019ae2,zhou2019dual}. \emph{Third}, many methods first extract features from multi-modal data, and then feed these features into a subsequent classification model (\emph{e.g.}, SVM). Due to the possible heterogeneity between the features and the model, ignoring the correlation between the two could lead to sub-optimal results. Intuitively, integrating both components into a unified framework could improve the prediction performance.

To tackle the above issues, we propose an end-to-end \textbf{Multi-modal Multi-channel Network} (\textbf{$\textup{M}^2\textup{Net}$}) for OS time prediction. $\textup{M}^2\textup{Net}$ first adopts a {modality-specific network} to automatically extract implicit and high-level features from different modalities (\emph{i.e.}, different MR scans), and then applies a {multi-modal shared network} to fuse them using a bilinear pooling model. Here, the modality-specific network is used to preserve the modality-specific attributes, while the multi-modal shared network is used to exploit the correlations. Further, we add another layer that automatically weights the outputs from each modality-specific network and the multi-modal shared network. Compared with conventional classification-based OS prediction methods, $\textup{M}^2\textup{Net}$ can not only extract high-level features to model complex tumor patterns, but also integrates feature learning and prediction model training into a unified framework. The unified framework enables label information to guide the feature learning more effectively. Besides, both the correlations across multi-modal data and the modality-specific properties can be seamlessly exploited. Finally, the experimental results demonstrate that our $\textup{M}^2\textup{Net}$ outperforms several classification-based OS prediction methods.

\section{Data Description and Processing}

The BraTS 2018 was organized using multi-institutional pre-operative MRI scans for the segmentation of intrinsically heterogeneous brain tumor sub-regions \cite{menze2015multimodal,havaei2017brain}. In this dataset, each patient includes 1) T1, 2) T1ce, 3) T2, and 4) FLAIR volumes. In this study, we focus on OS time prediction; thus, we have 163 subjects with survival information (in days) for this task.

\begin{figure*}[t]
	\captionsetup{font={small}}
	\begin{center}
		\includegraphics[width=0.85\textwidth]{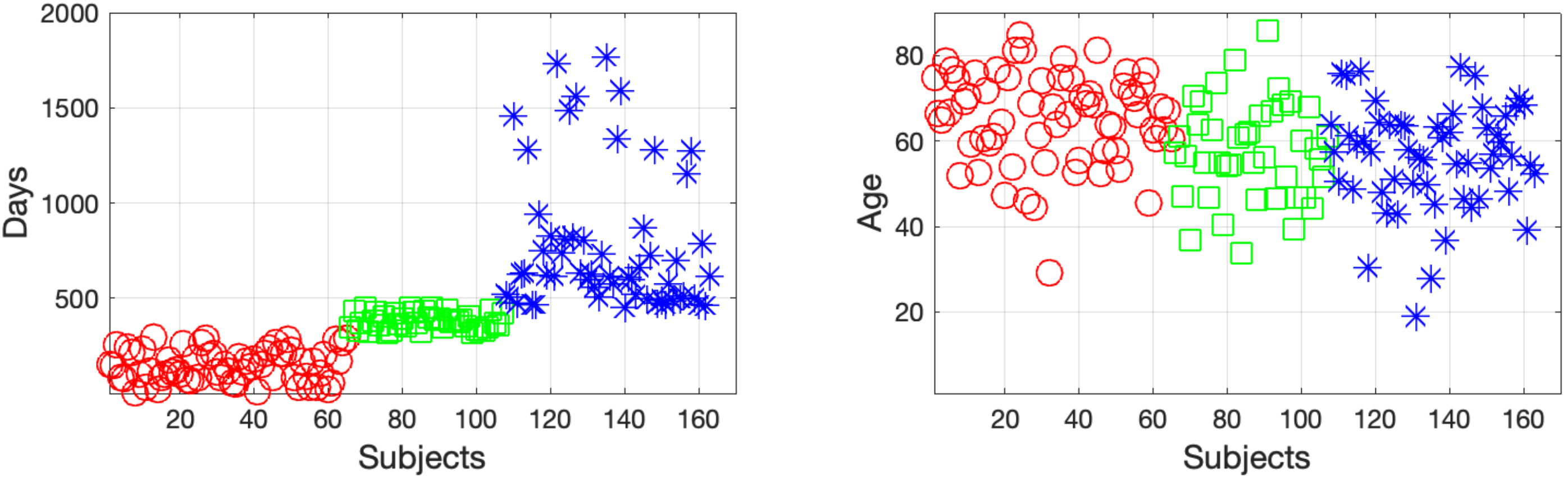}\vspace{-0.1cm}
		\caption{Illustration of subjects' survival time in days (left) and age information (right), where red, green and blue denote the information of short-term, mid-term, and long-term, respectively.}
		\label{fig0}
	\end{center}\vspace{-0.75cm}
\end{figure*}

\textbf{Definition of OS Time}. Similar to several previous studies \cite{zhou2017tpcnn}, we formulate the OS time prediction as classification-based task. Inspired by \cite{zhou2017tpcnn}, we define OS time as the duration from the date the patient was first scanned, to the date of the patient's tumor-related death. To construct a classification task, the continuous survival time can be divided into three classes: (1) short-term survivors (\emph{i.e.}, $\leq$10 months), (2) mid-term survivors (\emph{i.e.}, between 10 and 15 months), and (3) long-term survivors (\emph{i.e.}, $\geq$15 months). 
Fig.~\ref{fig0} shows subjects' survival time in days (left) and ages (right). As shown in Fig.~\ref{fig0},
it can clearly be seen that there are large intra-class differences, especially in the long-term class. Therefore, it is very challenging to achieve accurate prediction for OS time.

\textbf{Image Patch Extraction}. For our OS prediction task, we first locate the tumor region according to the tumor mask and extract an image patch that is centered at the tumor region. Then we resize the extracted image patch of each subject to a predefined size (\ie, $124\times124\times124$).

\textbf{Lesion and Age Information}. Because tumor size and age information may also be related to brain status, they can affect the prediction performance for OS time. Thus, we utilize these as supplemental features. Specifically, for every type of tumor, we calculate the size of each tumor using: $s_i=n_i/{\sum{n_i}}~(i=1,2,3)$, where $n_i$ is the number of the $i$-th type of tumor. There are three types of tumors, \emph{i.e.}, the necrotic and non-enhancing tumor core ($\text{label}=1$), the peritumoral edema ($\text{label}=2$), and the gadolinium-enhancing tumor ($\text{label}=4$). We also calculate the total size using $s_{total}=\sum{n_i}/n_{non-zero}$, where $n_{non-zero}$ is the total number of the non-zero elements in the original MR volume. Further, we add the age information (denoted as ``$s_{age}$") into the additional feature vector. Finally, we have the supplemental feature vector as $s=[s_{total},s_1,s_2,s_3,s_{age}]^{\top}$.

\section{The Proposed Method}

In this study, we effectively exploit the correlations among multi-modalities as well as preserve the model-specific attribute, for improving the OS prediction performance. We detail the three main parts: modality-specific network, multi-modal shared network, and multi-modal multi-channel network below.

\begin{figure*}[t]
	\captionsetup{font={small}}
	\begin{center}
		\includegraphics[width=0.9\textwidth]{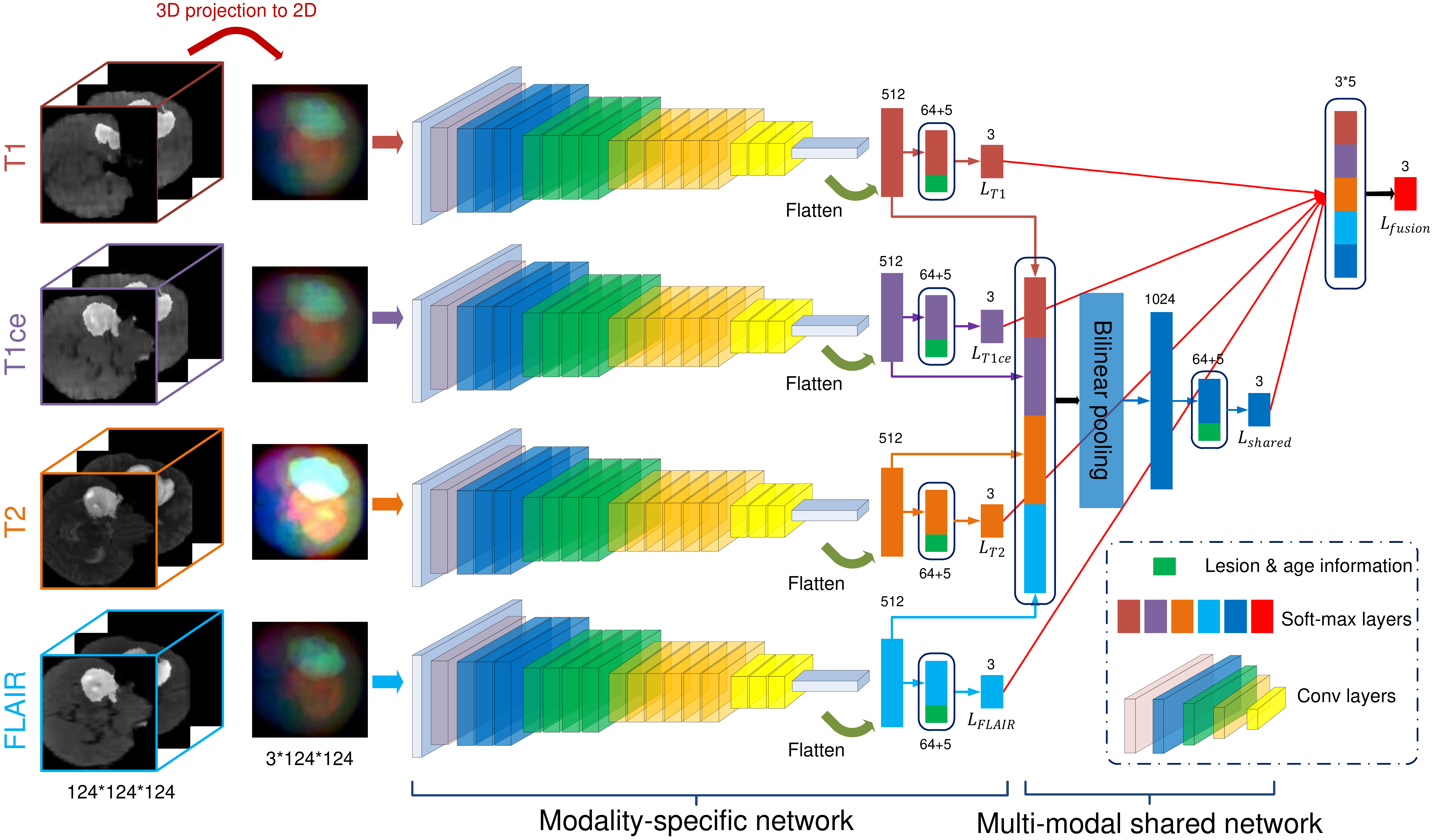}
		\caption{ Overview of the proposed $\textup{M}^2\textup{Net}$ for OS time prediction. \emph{First}, we project the 3D MR volumes onto three-channel 2D images in different directions. \emph{Second}, we propose an end-to-end OS time prediction model with multi-modal data.
		}
		\label{fig1}
	\end{center}\vspace{-0.45cm}
\end{figure*}

\vspace{-5pt}
\subsection{Modality-specific Network}

Convolutional neural networks (CNNs) are a class of deep neural networks that have been widely applied for analyzing visual tasks. Deep learning based CNN models can learn a hierarchy of features, in which high-level features are built upon low-level image features using a layer-wise strategy. Many pre-trained models have been developed and obtain very promising performances due to their being trained on large-scale datasets \cite{he2016deep}. In this study, in order to reduce computational costs while preserving important information and enabling a pre-trained model to be transferred from another related task, we project 3D MR volumes onto multi-channel 2D images, in different directions (as shown in Fig.~\ref{fig1}). Specifically, 3D images can be projected onto 2D images along each axis by averaging the sum of all slices from each volume. Using this strategy, we can obtain a three-channel 2D image. Then, we use the ResNet34 \cite{he2016deep} as the base network to extract modality-specific attributes from each modality (\emph{i.e.}, T1, T1ce, T2, and FLAIR). Note that, in ResNet34, the second last layer is a $7\times7$ max-pooling, which we change to a $4\times4$ max-pooling layer for our specific task. Thus, we have a fully-connected (FC) layer with $512$ nodes (while ResNet34 has $1024$ nodes in its FC layer). Finally, for each modality-specific network, we have the corresponding losses $\mathcal{L}_\text{T1}$, $\mathcal{L}_\text{T1ce}$, $\mathcal{L}_\text{T2}$, and $\mathcal{L}_\text{FLAIR}$, respectively.

\subsection{Multi-modal Shared Network}

Bilinear models \cite{lin2015bilinear} have been widely applied in fine-grained recognition, semantic segmentation, face recognition, and other fields, and have obtained impressive performance. Bilinear pooling often forms a descriptor by calculating
{\small$B(\mathcal{X})=\sum_{s\in\mathcal{S}}x_{s}x_{s}^{\top}$}, where {\small$\mathcal{X}=(x_1,\ldots,x_{|\mathcal{S}|};x_s\in\mathbb{R}^{c})$} denotes a set of local descriptors, and $\mathcal{S}$ denotes the set of spatial locations. Using this equation, we can obtain a $c\times{c}$ matrix. However, the classic bilinear features are often high-dimensional, making them impractical for many applications. To address this issue, a compact bilinear pooling method \cite{gao2016compact} has been developed to obtain the same discriminative ability as the full bilinear representation but with only few thousand dimensions. Further, given two sets of local descriptors, {\small$\mathcal{X}$} and {\small$\mathcal{Y}$}, a linear kernel machine can be used as follows:
\begin{equation}
\small
\begin{split}
\langle B(\mathcal{X}),B(\mathcal{Y})\rangle=\langle\sum_{s\in\mathcal{S}}x_{s}x_{s}^{\top},\sum_{u\in\mathcal{U}}y_{u}y_{u}^{\top}\rangle=\sum_{s\in\mathcal{S}}\sum_{u\in\mathcal{U}}\langle x_{s},y_{u}\rangle^{2},
\end{split}
\label{eq002-03}
\end{equation}
\noindent where $\mathcal{U}$ denotes another set of spatial locations.
If we could find some low dimensional projection function {\small$\varphi(x)\in\mathbb{R}^d$}, where {\small$d\ll{c}^2$}, that satisfies {\small$\langle\varphi(x),\varphi(y)\rangle\approx k(x,y)$}, we could approximate the inner product of (\ref{eq002-03}) by:
\begin{equation}
\small
\begin{split}
\langle B(\mathcal{X}),B(\mathcal{Y})\rangle=\sum_{s\in\mathcal{S}}\sum_{u\in\mathcal{U}}\langle x_{s},y_{u}\rangle^{2}\approx \sum_{s\in\mathcal{S}}\sum_{u\in\mathcal{U}}\langle \varphi(x),\varphi(y)\rangle^{2}=\langle{C}(\mathcal{X}),C(\mathcal{Y})\rangle,
\end{split}
\label{eq002-04}
\end{equation}
\noindent where {\small$C(\mathcal{X})=\sum_{s\in\mathcal{S}}\phi(x_s)$} is the compact bilinear feature. It is clear from this analysis that any low-dimensional approximation of the polynomial kernel can be used to obtain compact bilinear features.

In this study, to learn the compact discriminative representations for multi-modal data, we utilize the compact bilinear pooling method to effectively fuse them and exploit the correlations across multiple modalities (as shown in Fig.~\ref{fig1}). Thus, for the multi-modal shared network, we have the loss $\mathcal{L}_\text{shared}$.

\subsection{$\textup{M}^2\textup{Net}$: Multi-modal Multi-channel Network}

To effectively exploit the correlations among multi-modal data, as well as preserve modality-specific attributes for each modality, we propose a Multi-modal Multi-channel Network (\emph{i.e.}, $\textup{M}^2\textup{Net}$). As shown in Fig.~\ref{fig1}, we can obtain multiple outputs (\emph{i.e.}, $\text{out}_v^{sp},v=1,2,3,4$) for all modality-specific networks and one output (\emph{i.e.}, $\text{out}^{sh}$) from the multi-modal shared network. A direct way to fuse the various outputs would be to weight them, \emph{i.e.},  $\text{out}=\sum_{v}w_v^{sp}*\text{out}_v^{sp}+w^{sh}*\text{out}^{sh}$, where $w_v^{sp}$ and $w^{sh}$ denote the corresponding weights ($v=1,2,3,4$). However, in order to construct an end-to-end prediction model, we can cascade all the outputs (\emph{i.e.}, $\text{out}_v^{sp}$ and $\text{out}^{sh}$), and then add a soft-max layer after the cascaded layer. It is worth noting that this strategy  automatically weights the various outputs. More importantly, it enables our model to automatically learn the contributions of each modality and the multi-modal fusion. Subsequently, the total loss of our $\textup{M}^2\textup{Net}$ can be obtained as
\begin{equation}
\begin{split}
\mathcal{L}=\lambda_1(\mathcal{L}_\text{T1}+\mathcal{L}_\text{T1ce}+\mathcal{L}_\text{T2}+\mathcal{L}_\text{FLAIR}+\mathcal{L}_\text{shared})+\lambda_2\mathcal{L}_\text{fusion},
\end{split}
\label{eq03-04}
\end{equation}
\noindent where $\mathcal{L}_\text{fusion}$ denotes the loss of the final fusion for multiple outputs, and $\lambda_1$ and $\lambda_2$ are trade-off parameters.

\section{Experiments}

\subsection{Experimental Settings}
In our experiments, we adopt a $10$-fold cross-validation strategy for performance evaluation. We further split the training set into two groups as: training set ($80\%$) and validation set ($20\%$). The best model for the validation set is used to evaluate the testing set. We adopt four evaluation metrics, including accuracy, precision, recall, and F-score \cite{provost2001robust}. We calculate the precision and recall using a one-class-versus-all-other-classes strategy, and then calculate F-score using $\text{F-score}=\frac{2*\text{Precision}*\text{Recall}}{\text{Precision}+\text{Recall}}$. 
Finally, we report the mean and standard deviation.

We compare $\textup{M}^2\textup{Net}$ with the following methods: 
$\bullet$ \textbf{Baseline} method. Similar to the work \cite{nie2019multi}, we first extract features from the outputs of the first FC layer (\emph{i.e.}, with $512$ nodes) for each modality, and then cascade the features from the four modalities and the supplemental features into a single feature vector (\emph{i.e.}, with $4\times{512}+5$ dimension). Sparse learning~\cite{liu2009slep} is applied to the cascaded features, and then the selected features are fed to an SVM classifier. $\bullet$ \textbf{3D CNN fusion} method. We use a 3D CNN ~\cite{khvostikov20183d} for each modality (\emph{i.e.}, the extracted 3D patches) and then fuse them using a bilinear model (denoted as ``3D CNN + fusion"). $\bullet$ \textbf{TPCNN} \cite{zhou2017tpcnn}. This method uses a CNN model to extract features from multi-modal data and then employs XGBoost to build the regression model for OS time prediction. $\bullet$ \textbf{Multi-channel ResNet} method. We cascade the three-channel 2D images from all the four modalities into a $12\times{124}\times{124}$ image, and then we apply the pre-trained ResNet34 model for prediction. Here, a $1\times{1}$ convolutional layer must be added before the standard ResNet (since three-channel images are used as inputs). 
$\bullet$ \textbf{Modality-specific nets} method. This is a degraded version of our proposed model, which only uses the modality-specific network and then fuses all the outputs (\emph{i.e.}, soft-max layers) to obtain the final results. $\bullet$ \textbf{Multi-modal shared net} method. This is another degraded version of our proposed model, which employs the multi-modal shared network using a bilinear model and then obtains the final prediction result.
Note that, for fair comparison, we use the supplemental features for all comparison methods.

\noindent\textbf{Implementation Details}: The proposed $\textup{M}^2\textup{Net}$ is implemented using PyTorch, and the model is trained using Adam optimizer.
The decay rate is set to $10^{-4}$, and the maximum epoch is $100$. The batch size is set to $32$. Besides, 
the training data are augmented by vertical and horizontal flips, and rotation. The trade-off parameters are set to $\lambda_1=0.1$ and $\lambda_2=0.5$.

%


\subsection{Results Comparison}

Table~\ref{tab2} shows the prediction performance of all the comparison methods. As can clearly be seen, our proposed method performs consistently better than all the comparison methods in four metrics.
Compared with the baseline method, our method effectively improves the prediction performance. Note that the baseline method conducts feature learning and classification model learning using a two-steps strategy, while our method integrates them into an end-to-end framework. In contrast to the ``3D CNN fusion" method, our model first projects the 3D MR volume onto 2D images in different directions and then transfers a pre-trained model to the prediction task. The results clearly validate the effectiveness of this strategy. Our method is also shown to have a better prediction performance than the TPCNN method. Besides, it is worth noting that the ``Multi-channel ResNet" method obtains a even better performance than Baseline and ``3D CNN fusion" methods. Furthermore, the improvement of our model over ``Modality-specific nets" and ``Multi-modal shared net" clearly validates the importance of fusing multiple outputs from the modality-specific networks and the multi-modal shared network. Note that, while several other studies~\cite{feng2018brain} also investigate OS time prediction, they tend to primarily focus on tumor segmentation and then directly conduct a regression task, reporting a mean square error. Thus, it is not possible to directly compare our method with these methods. Overall, our method obtains a relatively better performance than other comparison methods.

\renewcommand\arraystretch{1.0}
\begin{table*}[t]
	\captionsetup{font={small}}
	\centering
	\begin{center}
		\footnotesize
		\caption{ Prediction results (mean $\pm$ standard deviation) of different methods. The best results are highlighted in \textbf{bold}.
		}\vspace{0.1cm}
		\begin{tabular}{p{3.3cm}|p{2.0cm}<{\centering}p{2.0cm}<{\centering}p{2.0cm}<{\centering}p{2.0cm}<{\centering}} 
			\hline
			Method     & Accuracy         & Precision        &Recall             & F-score             \\
			
			\hline

			Baseline~\cite{nie2019multi}              &$0.544 \pm 0.140$   &$0.478 \pm 0.177$  &$0.506 \pm 0.135$  &$0.489 \pm 0.152$ \\
			3D CNN fusion~\cite{khvostikov20183d}     &$0.587 \pm 0.093$   &$0.479 \pm 0.114$  &$0.533 \pm 0.071$  &$0.501 \pm 0.141$ \\
			TPCNN~\cite{zhou2017tpcnn}                                    &$0.636 \pm 0.000$   &- -                &- -                &- - \\
			Multi-channel ResNet                      &$0.624 \pm 0.084$   &$0.556 \pm 0.167$  &$0.572 \pm 0.091$  &$0.559 \pm 0.136$ \\
			Modality-specific nets                    &$0.621 \pm 0.069$   &$0.580 \pm 0.177$  &$0.576 \pm 0.067$  &$0.569 \pm 0.110$ \\
			Multi-modal shared net                    &$0.611 \pm 0.071$   &$0.449 \pm 0.071$  &$0.566 \pm 0.045$  &$0.499 \pm 0.056$ \\
			
			$\textup{M}^2\textup{Net}$                &$\textBF{0.664}\pm\textBF{0.061}$  &$\textBF{0.574} \pm \textBF{0.141}$  &$\textBF{0.613} \pm \textBF{0.075}$  &$\textBF{0.589} \pm \textBF{0.102}$ \\
			\hline
		\end{tabular}
		\label{tab2}
	\end{center}\vspace{-0.45cm}
\end{table*}

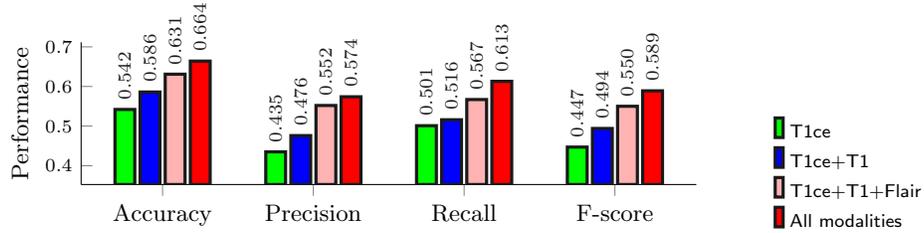
\begin{figure}[t]
	\begin{center}
		\captionsetup{font={small}}
		\begin{tikzpicture} 
		\begin{axis}[
		width=0.8\textwidth,
		height=0.18\textheight,
		ymin=0.4,
		ytick={0.4,0.5,0.6,0.7},
		ybar,
		enlargelimits=0.18,
		legend style={
			at={(1.25,0.52)},
			anchor=north,
			legend columns=1},
		legend cell align=left,
		ylabel={Performance},
		ylabel style={font=\footnotesize},
		yticklabel style = {font=\scriptsize},
		xticklabel style = {font=\footnotesize},
		legend style={draw=none},
		legend style={font=\scriptsize\sffamily},
		xtick=data,
		axis x line*=bottom,
		axis y line*=left,
		bar width=7.5pt,
		xticklabels={Accuracy, Precision, Recall, F-score},
		nodes near coords = \rotatebox{90}{{\pgfmathprintnumber[fixed zerofill, precision=3]{\pgfplotspointmeta}}}, 
		nodes near coords align={vertical}, 
		every node near coord/.append style={font=\scriptsize, yshift=0.01mm}, 
		]
		
		\addplot+[error bars/.cd, 
		y dir=both,y explicit]
		[color=black!90, very thick, fill=green!120, very thick]
		coordinates {(1,0.542) 
			(2,0.435) 
			(3,0.501) 
			(4,0.447)};
		
		\addplot+[error bars/.cd, 
		y dir=both,y explicit]
		[color=black!90, fill=blue!120, very thick]
		coordinates {(1,0.586) 
			(2,0.476) 
			(3,0.516) 
			(4,0.494) 	};
		
		\addplot+[error bars/.cd,  
		y dir=both,y explicit]
		[color=black!90, fill=pink!120, very thick]
		coordinates {(1,0.631) 
			(2,0.552) 
			(3,0.567) 
			(4,0.550) };
		
		\addplot+[error bars/.cd,  
		y dir=both,y explicit]
		[color=black!90, fill=red!120, very thick]
		coordinates {(1,0.664) 
			(2,0.574) 
			(3,0.613) 
			(4,0.589) };

		\legend{T1ce, T1ce+T1, T1ce+T1+Flair, All modalities}
		\end{axis} 
		\end{tikzpicture}\vspace{-0.3cm}
		\captionof{figure}{{Results comparison using multi-modal data fusion and single modality.}}\label{fig3}
	\end{center}
	\vspace{-0.65cm}
\end{figure}

\textbf{Multi-modal Data Fusion}. To analyze the effectiveness of the multi-modal data fusion, we first apply a modality-specific network for each modality, and we report the classification results for the best modality (\emph{i.e.}, T1ce in this experiment, as shown in Fig.~\ref{fig3}). Then, we conduct the experiment using a two-modality fusion, \emph{i.e.}, fusing T1ce and the three remaining modalities. We obtain the best performance when fusing the T1ce and T1 modalities. Next, we fuse the three modalities (\emph{i.e.}, T1ce+T1+FLAIR), and obtain an overall better performance that the two-modality fusion. From the results, it can be seen that the prediction performance of our method improves when using more modalities, which also verifies the effectiveness of multi-modal data fusion.

\vspace{-0.1cm}

\subsection{Discussion}

As reported in Table \ref{tab2}, our proposed model obtains much better prediction performance than all the comparison methods, the overall performance is still not very high. There could be several reasons for this. \emph{First}, for the OS time prediction task, over-fitting is very likely to occur since a very small dataset is used. We also note that the prediction task is different from the tumor segmentation task, because the segmentation task can be conducted on each slice from a 3D volume, providing significantly more samples.  \emph{Second}, as seen from Fig.~\ref{fig0}, there are large intra-class deviations in OS time, especially in the long-term class, and it is difficult to learn a reliable model to bridge these deviations. We also note that there is a small amount of subjects between short-term and middle-term classes that are very close together, which leads to ambiguous classification results.

In future work, we will focus on the exploration of features (including genetic and lesion information \cite{kao2018brain,tang2019pre}) through clinical collaboration, to improve the prediction performance. Additionally, modeling brain tumor growth is in itself meaningful, enabling us to better understand the mechanisms behind the disease progression and formulate better treatment planning. Thus, the mathematical modeling \cite{gevertz2006modeling,lipkova2019personalized} is a future direction. Besides, to deal with small sample issues, transfer learning \cite{li2018detecting,wang2016learning,zhou2020multi} can be introduced to our OS prediction task by borrowing prior knowledge from related tasks to improving prediction performance. We also use multi-stage fusion strategy \cite{zhou2019effective} to integrate multi-modal MR images for improving prediction performance. 
Further, we also use the generative adversarial networks \cite{MMsyn,fan2019adversarial,zhang2019skrgan,zhou2020hi} to synthesize more samples, which can be regarded as a form of data augmentation to enhance the prediction performance.

\section{Conclusion}
\label{conclusion}

In this paper, we propose a novel $\textup{M}^2\textup{Net}$ model to predict the OS time for brain tumor patients. Our $\textup{M}^2\textup{Net}$ model preserves the attributes of different modalities using a modality-specific network while also exploiting the correlations across multi-modal data using a multi-modal shared network. Further, the multiple outputs can be automatically fused to obtain the final prediction result. Experimental results demonstrate that our $\textup{M}^2\textup{Net}$ model improves OS time prediction accuracy, which also indicate the effectiveness of our proposed end-to-end prediction framework.

\noindent\textbf{Acknowledgement}. This research was supported in part by NSF of China (No: 61973090) and NSF of Tianjin (No: 19JCYBJC15200).

%
%
%

\bibliographystyle{splncs03}
\bibliography{refs}



\end{document}